\documentstyle[sprocl]{article}

\def \b{{\cal B}}
\def\s{\sqrt 2}

\def\be{\begin{equation}}
\def\ee{\end{equation}}
\def\bea{\begin{eqnarray}}
\def\eea{\end{eqnarray}}
\def\Dbar{\bar D^0}

\begin{document}

\rightline{TECHNION-PH-00-11}
\rightline{\large hep-ph/0001317}

\vskip 1.5cm
\title{DETERMINING THE CP VIOLATING PHASE $\gamma$
\footnote{To appear in Proceedings of The Third International Conference on 
B Physics and CP Violation, Taipei, December 3$-$7, 1999, H. -Y. Cheng and 
W. -S. Hou, eds. (World Scientific, 2000).}}

\author{MICHAEL GRONAU}

\address{\bigskip\bigskip \large
Department of Physics, Technion, Haifa 32000, Israel}

\maketitle\abstracts{\vskip 3cm \large
We review several methods for determining the 
Kobayashi-Maskawa phase $\gamma={\rm Arg} V^*_{ub}$ from rate and CP 
asymmetry measurements in hadronic $B$ decays. We focus on the processes 
$B\to DK,~B_s\to D_s K$, and on charmless decays to two light pseudoscalars 
and decays to a pair of a pseudoscalar and a vector meson. Theoretical 
uncertainties underlying these methods are discussed.}

\newpage
\section{Introduction}
$B$ meson decays open the window into new phenomena of CP nonconservation, 
providing useful information about CP violating phases \cite{MGrev}.
Phase measurements for $\alpha,~\beta$ and $\gamma$, which
are not independent in the Kobayashi-Maskawa framework, are important for two 
reasons. First, they improve to a higher 
precision our knowledge of the CKM mixing matrix, consisting of fundamental 
quark couplings in our present theory. And second, they are sensitive probes 
for sources of CP violation outside the CKM matrix, and for new flavor changing 
interactions contributing to $B^0-\bar B^0$ mixing and to rare $B$ decays. 

The phase $\beta \equiv{\rm Arg}(-V^*_{cb}V_{cd}/V^*_{tb}V_{td})~~ 
(={\rm Arg}V^*_{td}$ in the standard phase convention) is measured by the 
time-dependent CP asymmetry in $B^0(t)\to \psi K_S$ \cite{sanda}. Theoretically,
this measurement provides a very clean determination of $\sin(2\beta)$, since 
the single phase approximation holds in this case to better than 1$\%$ 
\cite{MG,LP}. 
The recent CDF measurement \cite{TRI}, $\sin(2\beta) = 0.79^{+0.41}_{-0.44}$, is 
an encouraging proof of the method, although it does not yet constitute 
definite evidence for CP violation in $B$ decays, as predicted in the CKM 
framework. 

The determination of $\sin 2\alpha$ ($\alpha\equiv{\rm Arg}(-V^*_{tb}
V_{td}/V^*_{ub}V_{ud})$) in $B^0\to\pi^+\pi^-$ suffers from the appearance of a 
second amplitude ($P$) due to QCD penguin 
operators \cite{MG,LP}. The weak phase \cite{Sinha} of $P$ differs from the 
phase of the dominant tree amplitude ($T$). The ratio 
$|P|/|T|=0.3\pm 0.1$, obtained \cite{DGR} by comparing within broken flavor 
SU(3) the measured rates \cite{CLEO} of 
$B\to\pi\pi$ and $B\to K\pi$, implies a potentially 
large deviation of the measured asymmetry from $\sin 2\alpha$. In fact, the 
time-dependent asymmetry in $B^0(t)\to\pi^+\pi^-$ contains two terms \cite{MG}
\be
A(t) = a_{\rm dir}\cos(\Delta mt) + \sqrt{1-a^2_{\rm dir}} 
\sin 2(\alpha + \theta)\sin(\Delta mt)~,
\ee
where $a_{\rm dir}$ and $\theta$ are both proportinal to $|P|/|T|$ and are 
functions of the relative strong phase between $P$ and $T$. A precise knowledge 
of $|P|/|T|$ would permit an accurate determination of $\alpha$ from a 
measurement of the two terms in the asymmetry \cite{MG,P/T}. One way of 
calculating $|P|/|T|$ in perurbative QCD is described by Deshpande at this 
meeting \cite{Desh}. However, the uncertainty in this calculation has not
yet been shown to be under control.

As long as $|P|/|T|$ cannot be calculated reliably, a theoretically cleaner way 
of resolving the penguin correction will require combining the asymmetry 
in $B^0\to\pi^+\pi^-$ with other measurements. A very early suggestion 
\cite{GL} was to also measure the rates of isospin related processes, 
$B^+\to\pi^+\pi^0$ and $B^0/\bar B^0\to \pi^0\pi^0$. However, it is expected
that the color-suppressed rates of tagged $B^0$ and $\bar B^0$ decays to 
neutral pions will involve large experimental errors, at least in the first 
round of experiments. Alternatively, $\alpha$ can be determined by applying the 
isospin technique to $B\to \rho\pi$ decays \cite{rhopi}, or by relating the 
time-dependence in $B^0(t)\to\pi^+\pi^-$ and in $B_s(t)\to K^+ K^-$ using flavor 
U-spin symmetry \cite{Bs}. Since none of these methods is expected to be both 
theoretically clean and experimentally accessible in the near future, some of 
these methods will have to be combined in order to reduce the error in $\alpha$.

The phase $\gamma\equiv{\rm Arg}(-V^*_{ub}V_{ud}/V^*_{cb}V_{cd})~
(={\rm Arg}V^*_{ub}$ in the standard convention)
is the relative weak phase between a CKM-favored 
($b\to c$) and a CKM-suppressed ($b\to u$) decay amplitude. Therefore, it 
contributes to 
direct CP asymmetries, in which two such amplitudes interfere, without the 
need for neutral $B$ mixing. Asymmetry measurements in charged $B$ decays and 
in self-tagged neutral $B$ decays have the advantage of 
avoiding the price of flavor tagging. However their theoretical interpretation
in terms of the weak phase $\gamma$ involves an unknown strong phase and an 
unknown ratio of two interferening hadronic matrix elements. A few methods were 
proposed to overcome this difficulty, often by measuring not only the 
asymmetry, but also the rates of certain related processes which provide 
information on these matrix elements. In the next few sections we describe 
several ways of measuring $\gamma$ in four classes of processes, 
discussing in each case both the theoretical uncertainties and the experimental 
limitations.

\section{$B\to DK^{(*)}$}

A simple idea for measuring $\gamma$ in $B^{\pm}\to D K^{\pm}$ was 
proposed some time ago \cite{GW}. Neglecting very small CP violation in $D^0-
\bar D^0$ mixing, one can write a triangle amplitude relation
\be
\sqrt{2}A(B^+\to D^0_1 K^+)=A(B^+\to D^0 K^+)+A(B^+\to \Dbar K^+)~.
\ee
$D^0_1$ is an even CP-eigenstate, decaying, for instance, to $\pi^+\pi^-$,
while $D^0$ and $\Dbar$ are two opposite flavor states. The two amplitudes 
on the right-hand-side contain CKM factors $V^*_{ub}V_{cs}$
and $V^*_{cb}V_{us}$, both of which are ${\cal O}(\lambda^3)$ \cite{Wolf},
and have a relative weak phase $\gamma$. If the two amplitudes were of 
about equal magnitudes, then the triangle relation and its 
charge-conjugate would permit a determination of $\sin\gamma$ within certain 
discrete ambiguities. 

At a closer examination one observes, however, that the first amplitude is 
suppressed relative to the second one by a CKM factor, 
$|V^*_{ub}V_{cs}|/|V^*_{cb}V_{us}|\approx 0.4$, and by a color-suppression 
factor of about 0.25. 
[This prediction comes from evidence for color-supression in $B\to \bar D\pi$ 
\cite{BHP}]. An order of magnitude suppression of this amplitude
relative to the measured amplitude \cite{DK-CLEO} of $B^+\to \Dbar K^+$, causes 
serious experimental difficulties in tagging the flavor of $D^0$, through decays 
such as $D^0\to K^-\pi^+$.
The very rare decay $B^+\to D^0K^+\to (K^-\pi^+)K^+$ interferes strongly with 
the doubly Cabibbo-suppressed decay of $\Dbar$ in $B^+\to\Dbar 
K^+\to 
(K^-\pi^+)K^+$. 

Two ways were proposed for partially evading this difficulty by considering 
only very rare decays, typically with branching ratios of order $10^{-7}$. Aside
from the small rates, both 
methods have other limitations.
\begin{itemize}
\item Study only color-suppressed decays \cite{GLD} $B^0\to D^0(\Dbar)K^{*0}$, 
where the flavors of neutral $D$ and $K^{*}$ are determined through 
$D^0\to K^-\pi^+,~K^{*0}\to K^+\pi^-$.  Here the interference between 
Cabibbo-favored and Cabibbo-suppressed neutral $D$ decays is much weaker.
Still, such interference occurs and prohibits a precise determination of 
$\gamma$.
\item Study $B^+\to D^0(\Dbar) K^+\to f K^+$, with two neutral $D$ decay modes, 
such as $f=K^-\pi^+,~K^-\rho^+$. In this case the two interfering amplitudes 
have comparable magnitudes, one being color-suppressed in $B$ decay and the 
other being doubly-Cabibbo-suppressed in $D$ decay. Measurement of the rates of 
these two processes and their charge-conjugates would permit a 
determination of $\gamma$~~\cite{ADS}, provided that the two doubly 
Cabibbo-suppressed 
$D$ decay branching ratios were known. Uncertainties in 
these branching ratios prohibit an accurate determination of $\gamma$.
An intrinsic uncertainty \cite{SISO} follows from the difficulty of 
disentangling doubly Cabibbo-suppressed $D^0$ decays from $D^0-\Dbar$ mixing
\cite{ASNER}.
The mixing may be larger than conventionally estimated \cite{DDbar},
as anticipated recently from large resonance contributions in 
$D^0$ decays \cite{res}. Therefore, a precise measurement of $\gamma$ using
this method requires knowledge of the $D^0-\Dbar$ mixing parameters. 
\end{itemize}

There are also variants \cite{Soffer}, which combine the above two schemes 
\cite{GW,ADS} based on CP and flavor states. In all cases, the very small 
branching ratios of the color-suppressed processes imply that such 
measurements cannot be carried out effectively in the first round of 
experiments at $e^+e^-$ $B$ factories, and would have to wait for hadron 
collider experiments providing at least $10^9$ $B$'s. 

An interesting question remains \cite{GDK}: What can be learned by studying 
only the more abundant color-allowed $B^{\pm}\to DK^{\pm}$ decays which have 
larger branching ratios? 
Considers the charge-averaged ratio of rates for neutral $D$ mesons 
decaying to an even (odd) CP state and for a color-allowed flavor state
\be\label{R_i}
R_i \equiv \frac{2[\Gamma(B^+ \to D_i K^+) + \Gamma(B^- \to D_i K^-)]}
{\Gamma(B^+ \to \Dbar K^+) + \Gamma(B^- \to D^0 K^-)}~~,~~~~~i = 1,2~~.
\ee
One finds (neglecting the small $D^0_1-D^0_2$ width-difference)
\be\label{R12}
R_{1,2} = 1 + \bar r^2 \pm 2\bar r\cos\bar\delta\cos\gamma~~,
\ee
where $A(B^+\to D^0 K^+)/A(B^+\to \Dbar K^+) \equiv \bar r\exp[i(\bar\delta+
\gamma)]$. 
This leads to two general inequalities
\be\label{LIMIT}
\sin^2\gamma \leq R_{1,2}~~,~~~~~i = 1,2~~,
\ee
one of which {\it must} imply a new constraint on $\gamma$, unless 
$\gamma=\pi/2$.

Assuming, for instance, $\bar r=0.1,~\bar\delta=0,~\gamma=40^\circ$, one finds 
$R_2=0.85$.
With $10^8~~B^+B^-$ pairs, and using measured $B$ and $D$ decay branching
ratios \cite{PDG}, one estimates an error \cite{GDK} $R_2=0.85 \pm 0.05$. In 
this case, Eq.(\ref{LIMIT}) excludes the range $73^\circ 
<\gamma < 107^\circ$ with 90$\%$ confidence level. Including measurements of the
CP asymmetries in $B^{\pm} \to D_i K^{\pm}$ permit, in principle, a determination
of $\gamma$ (and not only bounds on the angle). This depends, of course, on the 
unknown strong phase $\bar\delta$.
For further studies of related methods, see \cite{GRDK}.

\section{$B^0_s(t)\to D_s K$}

The interference between the two $\lambda^3$ subprocesses $b \to c\bar u s$ 
and $b\to u\bar c s$ operates also in the time-dependent decay $B^0_s\to 
D^-_s K^+$, leading to a $\sin(\Delta m_s t)$ term  \cite{tdep}
\bea\label{DsK}
\Gamma(B^0_s(t) \to D^-_sK^+)&=&e^{-\Gamma_st}[|A|^2\cos^2(\frac
{\Delta m_st}{2}) + |\bar A|^2\sin^2(\frac{\Delta m_st}{2})\nonumber\\
&-&|A\bar A|\sin(\delta+\gamma) \sin(\Delta m_s t)]~.
\eea
The two color-allowed amplitudes, corresponding to  $b \to c\bar u s$ and 
$b\to u\bar c s$, have magnitudes $|A|$ and $|\bar A|$, and involve relative 
strong and weak phases, $\delta$ and $\gamma$, respectively. 

These four parameters describe in a similar way
the time-dependence in the three processes in which the initial and/or 
final states are charge-conjugated, $\bar B^0_s(t) \to D^-_sK^+,~
B^0_s(t) \to D^+_sK^-,~\bar B^0_s(t) \to D^+_sK^-$.
Thus, measuring the time-dependent oscillations in these four processes, all 
of which require tagging the flavor of initial $B^0_s$, permits a 
determination of $\gamma$~~\cite{ADKD}. It is obvious that, for an accurate 
measurement, one would also need to know the width-difference
between the two $B_s$ mass eigenstates, which was neglected in (\ref{DsK}). 
Further studies and discussions of
width-dependent effects can be found in \cite{FLEIDU}.

\section{$B \to PP$}

We will consider $B$ decays to 
two light pseudoscalars, $B\to PP$ where $P=\pi, K$, within the framework of 
flavor SU(3) symmetry \cite{Zepp,Chau,GHLR,Grin}. Final states involving 
$\eta$ and $\eta'$ can be studied similarly \cite{eta}. Occasional reference
to SU(3) breaking effects will be made. A more ambitious approach, relying on
generalized factorization \cite{Desh,BBNS,Cheng,Li}, has not yet been justified 
quantitatively to a satisfactory level.

The low energy effective weak Hamiltonian \cite{BBL} governing $B\to PP$ 
\be\label{H}
{\cal H} = \frac{G_F}{\sqrt2}
\sum_{q=d,s}\left(\sum_{q'=u,c} \lambda_{q'}^{(q)}
[c_1 Q^{(q')}_1 + c_2 Q^{(q')}_2]
-  \lambda_t^{(q)}\sum_{i=3}^{10}c_i Q^{(q)}_i\right)~,
\ee
where $\lambda_{q'}^{(q)}=V_{q'b}^*V_{q'q}$, consists of the sum of three 
types of four quark operators: two $(V-A)(V-A)$ current-current operators 
($Q_{1,2}$), four 
QCD penguin operators ($Q_{3,4,5,6}$), and four electroweak penguin (EWP) 
operators ($Q_{7,8,9,10}$) with different chiral structures. The EWP operators 
with dominant Wilson coefficients, $Q_9$ and $Q_{10}~~(|c_{7,8}|/|c_9|
\approx 0.04$), have both a $(V-A)(V-A)$ structure.
All the four-quark operators, $(\bar bq_1)(\bar q_2 q_3)$, can be decomposed 
into a sum of ${\overline {\bf 15}}$, ${\bf 6}$ and ${\overline {\bf 3}}$ 
representations \cite{Zepp,GHLR,Grin}. The QCD penguin operators are pure 
${\overline {\bf 3}}$. 

A simple proportionality relation was noted recently \cite{EWPVP} to hold 
between the dominant EWP operators $Q_9$ and $Q_{10}$
and the current-current (``tree") operators $Q_1$ and $Q_2$, both trasforming 
as given SU(3) representations 

\bea\label{15}
{\cal H}^{(q)}_{EWP}(\overline{\bf 15}) &=& -\frac32 \frac{c_9+c_{10}}{c_1+c_2} 
\frac{\lambda_t^{(q)}}{\lambda_u^{(q)}} 
{\cal H}^{(q)}_T(\overline{\bf 15})~,\\
\label{6}
{\cal H}^{(q)}_{EWP}({\bf 6}) &=& \frac32 \frac{c_9-c_{10}}{c_1-c_2}
\frac{\lambda_t^{(q)}}{\lambda_u^{(q)}}{\cal H}^{(q)}_T({\bf 6})~.
\eea
The superscripts $q=d,s$ denote strangeness-conserving and 
strangeness-changing transitions, respectively. The two ratios of Wilson 
coefficients are equal within 3$\%$~~\cite{BBL}
\be
\frac{c_9+c_{10}}{c_1+c_2} \approx \frac{c_9-c_{10}}{c_1-c_2} \approx 
-1.12\alpha~.
\ee
As a consequence of (\ref{15}) and (\ref{6}), processes given by 
$\overline{\bf 15}$ and ${\bf 6}$ 
transitions contain EWP and current-current contributions which are 
proportional to each other. For instance, in the $\Delta S=1$
pure $\overline{\bf 15}$ transition, $B\to (K\pi)_{I=3/2}$, where 
$|I=3/2\rangle=|K^0\pi^+\rangle + \s|K^+\pi^0\rangle$, the ratio of these 
amplitudes 
\be\label{del}
\delta_{EW}=-\frac32 \frac{c_9+c_{10}}{c_1+c_2}\frac{|V^*_{tb}V_{ts}|}
{|V^*_{ub}V_{us}|}=0.65\pm 0.15 
\ee
is of order one \cite{Flei}, in spite of the fact that EWP amplitudes are 
higher order in electroweak couplings. 

This simple result, obtained in the limit of flavor SU(3),
was used \cite{NRPL} in order to obtain a bound on $\gamma$ in $B^{\pm}\to K\pi$.
Expressing $B^+\to K\pi$ in terms of reduced SU(3) amplitudes 
depicted in graphical form \cite{GHLR}, one has \cite{GPY}
\bea
A(B^+\to K^0\pi^+) &=&\lambda_t^{(s)}[P+EW] +|\lambda_u^{(s)}|e^{i\gamma}
[P_{uc}+A]~, 
\\
\s A(B^+\to K^+\pi^0)&=& -\lambda_t^{(s)}[P+EW]-|\lambda_u^{(s)}|[(T+C)
(e^{i\gamma}-\delta_{EW})+P_{uc}+A]~.\nonumber 
\eea
The dominant QCD penguin amplitudes $P$ in the two processes, carrying a weak
phase ${\rm Arg}\lambda_t^{(s)}=\pi$, are equal 
because of isospin. Unequal rates of the two processes would be evidence for
interference with smaller amplitudes involving a different weak phase. Defining 
the charge-averaged ratio of rates \cite{NRPL}
\be\label{R*}
R^{-1}_*\equiv \frac{2[B(B^+\to K^+\pi^0) + B( B^-\to K^-\pi^0)]}
{B(B^+\to K^0\pi^+) + B(B^-\to \bar K^0\pi^-)}~,
\ee
one finds, to leading order in small quantities
\be
R^{-1}_* = 1 - 2\epsilon \cos\phi (\cos\gamma - \delta_{EW}) + 
{\cal O}(\epsilon^2) + {\cal O}(\epsilon\epsilon_A) +{\cal O}(\epsilon^2_A)~.
\ee
$\epsilon$ and $\epsilon_A$ are ratios of tree-to-penguin and 
rescattering\cite{rescat} -to-penguin amplitudes, respectively
\be
\epsilon e^{i\phi}=\frac{|\lambda_u^{(s)}|}{|\lambda_t^{(s)}|}\frac{T+C}{P+EW}~,
~~~\epsilon_A=\frac{|\lambda_u^{(s)}[P_{uc}+A]|}{|\lambda_t^{(s)}[P+EW]|}~.
\ee
The first ratio is given by \cite{CLEO,GRL}
\be
\epsilon=\sqrt2 \frac{V_{us}}{V_{ud}}\frac{f_K}{f_\pi}
\frac{|A(B^+\to \pi^0\pi^+)|}{|A(B^+\to K^0\pi^+)|} = 0.21\pm 0.05~,
\ee
while the second one is roughly \cite{CLEO,falk}
\be
\epsilon_A \sim \lambda\frac{|A(B^+\to \bar K^0 K^+)|}{|A(B^+\to K^0\pi^+)|}
< 0.12
\ee

Neglecting second order terms, one obtains a bound \cite{NRPL}
\be\label{bound}
|\cos\gamma - \delta_{EW}| \ge \frac{|1-R^{-1}_*(K\pi)|}{2\epsilon}~,
\ee
which would provide useful information about $\gamma$ if a 
value different from one were measured for $R^{-1}_*$. The present value, 
$R^{-1}_{*~\rm exp}=1.27\pm 0.48$, is consistent with one.
Further information about $\gamma$, applying also to the case $R^{-1}_{*}=1$, 
can be obtained by measuring separately $B^+$ and $B^-$ decay rates \cite{NRPRL}.
The solution obtained for $\gamma$ involves uncertainties due to SU(3) breaking
in subdominant amplitudes and an uncertainty in $|V_{ub}/V_{cb}|$, both of which 
affect the value of $\delta_{EW}$. Combined with errors in 
$\epsilon \propto |A(B^+\to\pi^+\pi^0)/A(B^+\to K^0\pi^+)|$, and in the 
rescattering parameter $\epsilon_A$, the resulting uncertainty in $\gamma$ is 
unlikely to be smaller than $20^\circ$ \cite{NRPRL}. For other ways of studying 
$\gamma$ in $B\to K\pi$, see \cite{FLE}.

\section{$B \to VP$}

$B$ mesons decays to a charmless vector meson ($V$) and a pseudoscalar meson 
($P$) involve a larger number of SU(3) amplitudes \cite{vp1,vp2} than $B\to PP$.
SU(3) relations between EWP and current-current amplitudes, following from
(\ref{15}) and (\ref{6}), reduce considerably the number of independent
amplitudes. In this section we describe briefly two applications \cite{EWPVP}.

\subsection{$B^{\pm}\to \rho K$}

Defining a charge-averaged ratio of rates for $B^{\pm}\to\rho K$
\be
R^{-1}_*(\rho K)\equiv \frac{2[B(B^+\to \rho^0 K^+) + B( B^-\to \rho^0 K^-)]}
{B(B^+\to \rho^+K^0) + B(B^-\to \rho^-\bar K^0)}~,
\ee
one obtains, with some analogy to (\ref{bound}), the bound
\be
|\cos\gamma| \ge \frac{|1-R^{-1}_*(\rho K)|}{2\epsilon_V} - \delta_{EW}\left(
\frac{\epsilon_P}{\epsilon_V}\right )~,
\ee
where
\be
\epsilon_V = \sqrt2 \frac{V_{us}}{V_{ud}}\frac{f_K}{f_\pi}
\frac{|A(B^+\to \rho^0\pi^+)|}{|A(B^+\to \rho^+K^0)|},~~
\epsilon_P = \sqrt2 \frac{V_{us}}{V_{ud}}\frac{f_{K^*}}{f_\rho}
\frac{|A(B^+\to \rho^+\pi^0)|}{|A(B^+\to \rho^+K^0)|}.
\ee
Although this constraint is weaker than (\ref{bound}), it shows that
measuring charge-averaged ratios of rates, which differ from one, is of 
interest also for $PV$ final states.

\subsection{$B^0\to K^{*\pm}\pi^{\mp}$ vs. $B^{\pm}\to\phi K^{\pm}$}

Considering the amplitudes for $B^0\to K^{*+}\pi^-$ and $B^+\to \phi K^+$, both
of which are expected to be QCD penguin dominated, and
keeping only dominant and subdominant terms, one finds
\bea
A(B^0\to K^{*+}\pi^-) &=& -\lambda^{(s)}_t\,P_P - \lambda^{(s)}_u\,T_P~,
\nonumber\\ 
A(B^+\to \phi K^+) &=& \lambda^{(s)}_t\,[P_P+EW]~,
\eea
where the suffix $P$ denotes the pseudoscalar which includes the spectator in
the graphic SU(3) amplitudes \cite{vp1}. The EWP contribution is related to the 
tree amplitude by \cite{EWPVP}
\be
\lambda^{(s)}_t EW = \frac{1}{3}\delta_{EW}\,|\lambda^{(s)}_u|T_P~.
\ee
Defining the ratio
\be
R=\frac{|A(B^0\to K^{*+}\pi^-)|^2+|A(\bar B^0\to K^{*-}\pi^+)|^2}
{|A(B^+\to \phi K^+)|^2+|A(B^-\to \phi K^-)|^2}~,
\ee
one finds 
\be
R=\frac{1+r^2-2r\,\cos\delta\cos\gamma}{1+(\delta_{EW}/3)^2\,r^2-(2/3)\,
\delta_{EW}\,r\,\cos\delta}~,
\ee
where 
\be
r\,e^{i\delta} = \frac{|\lambda^{(s)}_u|}{|\lambda^{(s)}_t|}\frac{T_P}{P_P}~.
\ee

Present 90$\%$ confidence level limits \cite{CLEO}, 
$\b(B^0\to K^{*\pm}\pi^{\mp})>12\times 10^{-6}$ and $\b(B^{\pm}\to \phi 
K^{\pm})<5.9\times 10^{-6}$, imply $R>2$, which is evidence for a nonzero 
contribution of $T_P$. In order to use this inequality for information about 
$\gamma$, 
one must include some input about $r$ and $\delta$.
A reasonable assumption, supported both by perturbative \cite{BBNS} and 
statistical \cite{Suz} calculations, is that $\delta$ does not exceed $90^\circ$. 
A very conservative assumption about $r$ is $r\le 1$. Making these two
assumptions, one finds [for $\delta_{EW}$ we will use the range (\ref{del})]
\be
\cos\gamma -\frac{2}{3}\delta_{EW} < \frac{-1 + r^2[1-2(\delta_{EW}/3)^2]}{2r}~.
\ee
This implies $\gamma > 62^\circ$ for $r=1$, and $\gamma > 105^\circ$ for 
$r=0.5$. 
Indirect evidence \cite{vp2} exists already for $r<0.55$. Direct 
information on $r$ will be obtained from future rate 
measurements of $B^+\to K^{*0}\pi^+$ and $B^+\to\rho^+\pi^0$ 
or $B^0\to\rho^+\pi^-$.

These bounds neglect smaller terms in the amplitudes, primarily the 
color-suppressed tree amplitude. If this contribution is $10\%$ ($20\%$) of 
the color-favored tree amplitude $T_P$ \cite{BBNS}, then
the limits move up or down by about $5^\circ~(10^\circ)$. The bounds also 
assume [by SU(3)] equal QCD penguin contributions in the two processes. 
The constraint becomes stronger if the penguin amplitude in 
$B^+\to \phi K^+$ is larger than in $B^0\to K^{*+}\pi^-$, as predicted by 
factorization \cite{ALI}. However, the constraint may become weaker if SU(3) breaking
in penguin amplitudes is not described by factorization.

\section{Conclusion}

We discussed several ways of determining the weak phase $\gamma$. The first 
two schemes, based on $B\to DK$ and $B_s\to D_s K$, seem at first sight to 
involve no hadronic uncertainties. However,
at a closer look, they require taking care of smaller effects,
such as $D^0-\Dbar$ mixing and the width-difference in the $B^0_s$ system.
Methods based on charmless decays to two pseudoscalars, and decays to a 
pseudoscalar and a vector meson, involve dynamical hadronic parameters,
such as those describing SU(3) breaking and rescattering amplitudes. Some of 
these quantities are under control, and others should be studied through a 
dialogue between theory and experiments.

\section*{Acknowledgments}
This work was supported in part by the Israel Science Foundation founded by the 
Israel Academy of Sciences and Humanities, by the United States -- Israel 
Binational Science Foundation under Research Grant Agreement 98-00237, by 
the Technion V.P.R. fund - Harry Werksman Research Fund, and by P. and E. 
Natahan Research Fund. 
\bigskip

\def \ajp#1#2#3{Am. J. Phys. {\bf#1}, #2 (#3)}
\def \apny#1#2#3{Ann. Phys. (N.Y.) {\bf#1}, #2 (#3)}
\def \app#1#2#3{Acta Phys. Polonica {\bf#1}, #2 (#3)}
\def \arnps#1#2#3{Ann. Rev. Nucl. Part. Sci. {\bf#1}, #2 (#3)}
\def \art{and references therein}
\def \cmts#1#2#3{Comments on Nucl. Part. Phys. {\bf#1}, #2 (#3)}
\def \cn{Collaboration}
\def \cp89{{\it CP Violation,} edited by C. Jarlskog (World Scientific,
Singapore, 1989)}
\def \epjc#1#2#3{Euro.~Phys.~J.~C {\bf #1}, #2 (#3)}
\def \epl#1#2#3{Europhys.~Lett.~{\bf #1}, #2 (#3)}
\def \ib{{\it ibid.}~}
\def \ibj#1#2#3{~{\bf#1}, #2 (#3)}
\def \ijmpa#1#2#3{Int. J. Mod. Phys. A {\bf#1}, #2 (#3)}
\def \jpb#1#2#3{J.~Phys.~B~{\bf#1}, #2 (#3)}
\def \jhep#1#2#3{JHEP {\bf#1}, #2 (#3)}
\def \mpla#1#2#3{Mod. Phys. Lett. A {\bf#1}, #2 (#3)}
\def \nc#1#2#3{Nuovo Cim. {\bf#1}, #2 (#3)}
\def \np#1#2#3{Nucl. Phys. {\bf#1}, #2 (#3)}
\def \pisma#1#2#3#4{Pis'ma Zh. Eksp. Teor. Fiz. {\bf#1}, #2 (#3) [JETP Lett.
{\bf#1}, #4 (#3)]}
\def \pl#1#2#3{Phys. Lett. {\bf#1}, #2 (#3)}
\def \pla#1#2#3{Phys. Lett. A {\bf#1}, #2 (#3)}
\def \plb#1#2#3{Phys. Lett. B {\bf#1}, #2 (#3)}
\def \pr#1#2#3{Phys. Rev. {\bf#1}, #2 (#3)}
\def \prc#1#2#3{Phys. Rev. C {\bf#1}, #2 (#3)}
\def \prd#1#2#3{Phys. Rev. D {\bf#1}, #2 (#3)}
\def \prl#1#2#3{Phys. Rev. Lett. {\bf#1}, #2 (#3)}
\def \prp#1#2#3{Phys. Rep. {\bf#1}, #2 (#3)}
\def \ptp#1#2#3{Prog. Theor. Phys. {\bf#1}, #2 (#3)}
\def \ptwaw{Plenary talk, XXVIII International Conference on High Energy
Physics, Warsaw, July 25--31, 1996}
\def \rmp#1#2#3{Rev. Mod. Phys. {\bf#1}, #2 (#3)}
\def \rp#1{~~~~~\ldots\ldots{\rm rp~}{#1}~~~~~}
\def \stone{{\it $B$ Decays} (Revised 2nd Edition), edited by S. Stone
(World Scientific, Singapore, 1994)}
\def \yaf#1#2#3#4{Yad. Fiz. {\bf#1}, #2 (#3) [Sov. J. Nucl. Phys. {\bf #1},
#4 (#3)]}
\def \zhetf#1#2#3#4#5#6{Zh. Eksp. Teor. Fiz. {\bf #1}, #2 (#3) [Sov. Phys. -
JETP {\bf #4}, #5 (#6)]}
\def \zpc#1#2#3{Zeit. Phys. C {\bf#1}, #2 (#3)}
\def \zpd#1#2#3{Zeit. Phys. D {\bf#1}, #2 (#3)}

\end{document}